\documentclass[runningheads]{llncs}

\usepackage[T1]{fontenc}

\usepackage{graphicx}
\usepackage{subcaption}
\usepackage{tabularray}
\usepackage{array}
\usepackage{tabularx}
\usepackage{ragged2e}
\usepackage{booktabs}
\usepackage{float}
\usepackage{amsmath}
\PassOptionsToPackage{hyphens}{url}
\usepackage{hyperref}
\usepackage[capitalise,noabbrev,nameinlink]{cleveref}
\usepackage{xspace}
\usepackage{siunitx}
\usepackage{pifont}
\usepackage{threeparttable}
\usepackage{orcidlink}
\hypersetup{
    colorlinks=true,
    linkcolor=blue,
    filecolor=blue,
    urlcolor=blue,
    citecolor=blue,
    pdfborder={0 0 0}
}
\newcommand{\codename}{NovaClass\xspace}

\newcounter{principle}
\renewcommand{\theprinciple}{\arabic{principle}}

\newenvironment{principle}[1][]{
  \refstepcounter{principle}
  \par\medskip\noindent\textbf{P\theprinciple\quad#1}\par
}{\medskip}

\crefformat{principle}{#2P#1#3}
\crefrangeformat{principle}{#3P#1#4–#5P#2#6}
\crefmultiformat{principle}{#2P#1#3}{ and #2P#1#3}{, #2P#1#3}{, and #2P#1#3}

\begin{document}

\title{Engineering Trustworthy Automation:\\ Design Principles and Evaluation for\\ AutoML Tools for Novices}
\titlerunning{Engineering Trustworthy AutoML Tools for Novices}

\author{Jarne Thys\,\orcidlink{0009-0005-9348-7405} \and
Davy Vanacken\,\orcidlink{0000-0001-8436-5119} \and
Gustavo Rovelo Ruiz\,\orcidlink{0000-0001-7580-8950}
}

\authorrunning{J. Thys et al.}

\institute{UHasselt - Hasselt University, Digital Future Lab - Flanders Make
\email{\href{mailto:jarne.thys@uhasselt.be}{jarne.thys@uhasselt.be}}
\email{\href{mailto:davy.vanacken@uhasselt.be}{davy.vanacken@uhasselt.be}}
\email{\href{mailto:gustavo.roveloruiz@uhasselt.be}{gustavo.roveloruiz@uhasselt.be}}
}

\maketitle

\begin{abstract}
AutoML systems targeting novices often prioritize algorithmic automation over usability, leaving gaps in users' understanding, trust, and end-to-end workflow support. To address these issues, we propose an abstract pipeline that covers data intake, guided configuration, training, evaluation, and inference. To examine the abstract pipeline, we report a user study where we assess trust, understandability, and UX of a prototype implementation. In a 24-participant study, all participants successfully built their own models, UEQ ratings were positive, yet experienced users reported higher trust and understanding than novices. Based on this study, we propose four design principles to improve the design of AutoML systems targeting novices: (P1) support first-model success to enhance user self-efficacy, (P2) provide explanations to help users form correct mental models and develop appropriate levels of reliance, (P3) provide abstractions and context-aware assistance to keep users in their zone of proximal development, and (P4) ensure predictability and safeguards to strengthen users' sense of control.

\keywords{AutoML \and Large Language Models \and Transformers \and Text Classification \and Conversational Assistant.}
\end{abstract}

\section{Introduction}

Novices are increasingly interested in training AI models, both professionally and personally. Professionally, AI's growing impact across industries creates a need for workers to develop AI competencies to remain competitive in evolving job markets~\cite{brawner_teaching_2023}. On a personal level, many novices are motivated by a genuine interest in AI technology itself. Some participants in studies expressed pride in contributing to the improvement of AI, highlighting the emotional connection and personal interest in AI development~\cite{you_labeling_2022}. Additionally, the potential for complementary performance between humans and AI systems motivates engagement. When humans and AI collaborate effectively, their combined performance can exceed what either could achieve alone, a concept known as complementary team performance~\cite{hemmer_effect_2022}. This potential becomes especially relevant in organizational settings where AI systems are deployed alongside subject matter experts, not only to support task execution but also to help transfer expert knowledge to less experienced users~\cite{spitzer_training_2022}. As a result, such collaboration provides a compelling reason for novices to invest in developing AI competencies, both to enhance their own performance and to contribute meaningfully to human-AI teams.

At the same time, new advanced AI/ML architectures can be applied to the specialized domains of those users. A professor could use a Transformer-based classifier to automatically grade open-ended questions, a chemist could use Graph Neural Networks to simulate reactions using molecular graphs, and clinicians could model continuous physical processes (e.g., ECG, EEG, EMG) using Liquid Neural Networks. However, despite the technological progress, the practical adoption of these advanced architectures by AI/ML novices remains limited. Deploying advanced architectures typically requires fluency with programming, familiarity with AI/ML frameworks, and careful attention to error-prone details such as dataset validation, model configuration, and training orchestration. As a result, domain experts without AI/ML expertise often struggle to leverage these models for their own data~\cite{paleyes_challenges_2022}.

Efforts to lower these barriers include AutoML platforms and no-code/low-code AI systems, which automate individual steps of the pipeline and expose model training through graphical interfaces. While such tools improve accessibility, they continue to leave critical gaps for novices. First, they often emphasize optimization of algorithms while offering little support for end-to-end workflows such as dataset validation, feature selection, and inference setup~\cite{xin_whither_2021}. Second, they rarely evaluate how their abstractions affect user trust and interpretability, focusing on the technical aspect and leaving open questions about how novices actually experience such systems~\cite{lindauer_position_2024}.

To address these challenges, our contributions are threefold:

\begin{enumerate}
    \item \textbf{Abstract AutoML Pipeline for Novices.} We propose an abstract end-to-end pipeline designed to support novices that links data intake, configuration, training, evaluation, and inference.
    \item \textbf{Evaluation of the Pipeline via a Prototype for Novice-Oriented Workloads.} A 24-participant study tests (i) end-to-end feasibility across various datasets and tasks using Transformer-based text classification, and (ii) robustness of the training and inference via metadata-driven pipelines.
    \item \textbf{Design Principles for AutoML Tools for Novices} Based on the study results and relevant theories, we present four design principles for future AutoML systems.
\end{enumerate}

\section{Related Work}

We review three topics relevant to AutoML tools for novices. First, we consider AutoML solutions that automate model selection and optimization, noting how they often leave workflow orchestration and user understanding to the side. Second, we examine no- and low-code platforms that broaden access but can obscure key decisions or provide limited support for diagnosis. Third, we look at AI assistants and contextual help that promise step-aware guidance but vary in reliability and integration.

\paragraph{\textbf{AutoML Solutions and Challenges.}}

AutoML is an alternative for users with limited technical knowledge. These platforms automate various components of the machine learning workflow to lower the entry barrier for novices. Notable systems include Auto-sklearn, TPOT, and commercial platforms like Google AutoML, DataRobot, and Azure Machine Learning Studio~\cite{olson_evaluation_2016,smith_machine_2020}. These systems employ different search strategies, such as Bayesian optimization combined with meta-learning~\cite{snoek_practical_2012}, genetic programming~\cite{olson_evaluation_2016}, and reinforcement learning~\cite{drori_alphad3m_2021}, to automatically generate and optimize ML pipelines. However, AutoML platforms face several limitations that prevent them from achieving their promise of fully automated machine learning. First, much of AutoML research has focused on isolated parts of the ML pipeline, such as preprocessing or hyperparameter optimization, rather than full end-to-end workflows, which often makes these methods difficult to apply without expert oversight ~\cite{smith_machine_2020}. Second, the search process for high-performance models can be extremely slow, taking minutes to hours, which affects system interactivity and necessitates asynchronous communication channels~\cite{egele_asynchronous_2023}. Consequently, studies show that AutoML users are still primarily expert data scientists, and the tools require skilled users~\cite{crisan_fits_2021,feng_addressing_2023}.

\paragraph{\textbf{No-Code/Low-Code AI Platforms.}}

To cater to a non-expert audience, different no-code and low-code platforms have been developed. These platforms represent the most accessible tier of interactive machine learning systems, designed to enable users with little to no programming expertise to create and deploy AI models through graphical interfaces~\cite{mumuni_automated_2025}. They typically provide user-friendly interfaces with tools that can automatically handle data processing tasks, such as finding missing data, identifying incorrect labels, and selecting desired data subsets~\cite{mumuni_automated_2025}. Recent advances leverage Large Language Models (LLMs) to create conversational interfaces that can iteratively extract user requirements and provide real-time guidance throughout the model-building process~\cite{luo_autom3l_2024,tayebi_arasteh_large_2024}. These natural language interfaces show particular promise for bridging technical knowledge gaps, allowing users across different expertise levels to successfully complete complex machine learning tasks~\cite{yao_evaluation_2025}.

\paragraph{\textbf{AI Assistants in Complex Software Systems.}}

The use of AI assistants in complex systems addresses a key limitation: while current techniques handle simple tasks effectively, they struggle to generalize to conversational interfaces that help humans solve complex problems through interaction with AI reasoning systems~\cite{allen_conversational_2020}. In practice, this spans domains from software engineering, where assistants ask clarifying questions and generate code~\cite{ross_programmers_2023}, to business operations, where intelligent task assistants execute processes via multi-agent orchestration that maps natural-language requests to executable sequences of operations connected to back-end services~\cite{chakraborti_d3ba_2020,he_rebalancing_2023}. These Conversational Agentic Systems combine the conversational capabilities of LLMs with structured function calls and typically require specialized dialogue fine-tuning to preserve coherence over extended workflows~\cite{robino_conversation_2025}. Such assistants can be further extended by taking context into account when providing support.

\paragraph{\textbf{Contextual Help and Guidance Systems.}}

The effectiveness of contextual help systems depends heavily on their ability to predict and respond to user satisfaction and engagement in real-time, particularly in open-domain conversations without clearly defined goals~\cite{choi_offline_2019}. Conversational interfaces offer notable advantages over traditional WIMP (Windows, Icons, Menus, and Pointers) interfaces by providing natural and familiar interaction methods, flexible accommodation of diverse user requests, and anthropomorphic features that help attract attention and gain trust, yet they continue to face significant challenges in processing natural language expressions and managing complex conversation situations~\cite{grudin_chatbots_2019,xiao_tell_2020}. LLMs can be used to address the limitations of previous systems in complex conversations. They have significantly enhanced contextual guidance capabilities, with multi-turn conversational prompting making LLMs more responsive and proficient in handling complex queries and extended discussions. These systems now demonstrate improved fluidity and relevance in interactions, making them more engaging and helpful across applications ranging from customer service to therapy bots~\cite{ding_data_2024}.

\section{Abstract Pipeline for AutoML Tools for Novices}

\begin{figure}
    \centering
    \includegraphics[width=\linewidth]{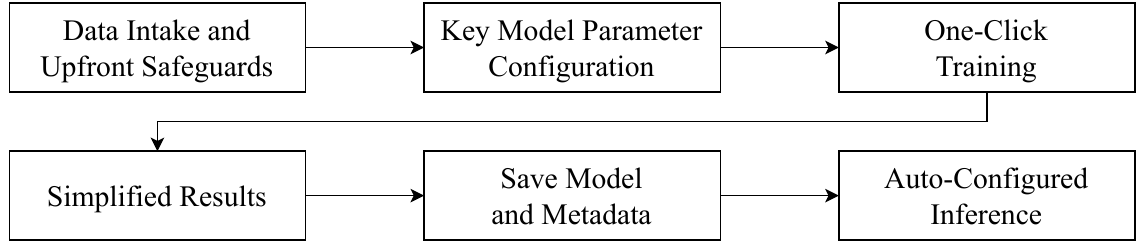}
    \caption{An abstract, end-to-end pipeline for AutoML tools targeting novice users. It begins with Data Intake and Upfront Safeguards, then narrows decisions to Key Model Parameter Configuration before One-Click Training. After training, Simplified Results explain performance in simplified terms, while Save Model and Metadata preserves an inference ``contract.'' Finally, Auto-Configured Inference uses that contract to generate input fields and outputs consistently.}
    \label{fig:abstract-diagram}
\end{figure}

Tools that create abstractions from advanced AI/ML architectures (e.g., Transformers, Graph Neural Networks, Liquid Neural Networks) can empower non-experts, but only if the abstraction is designed for usability and understanding; automation alone is not enough~\cite{lindauer_position_2024}. In this section, we introduce an abstract pipeline of an AutoML tool to support novices. The full pipeline is illustrated in \cref{fig:abstract-diagram}.

\begin{description}

\item[Data Intake and Upfront Safeguards.]{The pipeline begins with data ingestion and validation mechanisms. Automated data processing and feature engineering can streamline data preparation tasks~\cite{mumuni_automated_2025}. Domain experts can contribute knowledge relevant to data preprocessing and feature engineering~\cite{lindauer_position_2024}, safeguards at this stage address data quality and potential biases~\cite{crisan_fits_2021}.}

\item[Key Model Parameter Configuration.]{The user should only be exposed to parameters that fundamentally change the model and where their domain knowledge is relevant~\cite{lindauer_position_2024}, such as the selection of input features and the objective of the model. Other parameters (e.g., model, optimizer, learning rate) can either be auto-configured by the tool or be a static, robust baseline that can deliver sufficient performance for most cases. We argue that peak performance is not a requirement for experimentation.}

\item[One-Click Training.]{After the key parameters have been configured, the tool should handle all configuration without any user input. All parts of the ML pipeline (e.g., train-val-test splitting, handling missing values, converting string labels to integers) should be handled automatically.}

\item[Simplified Results.]{Transparency and interpretability affect user trust in AutoML tools, with interpretability identified as a key user requirement~\cite{lindauer_position_2024}. Result presentation should accommodate different stakeholder needs, from domain experts to ML practitioners~\cite{feng_addressing_2023,yao_evaluation_2025}.}

\item[Save Model and Metadata.]{This stage addresses model checkpointing, saving, and inference through metadata. All relevant training data that can be carried over to inference should be saved, as this metadata can be used as type hints for the user to correctly use the trained model and for the tool to configure a trained model for inference without any user input.}

\item[Auto-Configured Inference.]{The final stage provides automated deployment with provisions for human oversight in high-stakes applications. Inference should automate technical aspects of model serving~\cite{lindauer_position_2024} based on the metadata saved in the previous stage. The design should combine human expertise and AutoML capabilities, particularly for strategic decisions, ethical considerations, and domain-specific requirements~\cite{crisan_fits_2021,lindauer_position_2024}.}

\end{description}

\section{\codename: Applying the Abstract Pipeline to Transformer-based Classification}

To put the abstract pipeline into practice, the following section details \codename, our novice-friendly automation prototype, which aims to lower the entry barrier for novices who wish to fine-tune Transformer models for text classification tasks. We use supervised text classification as it has many use cases (e.g., grading, spam detection, emotion classification), and performance has taken a leap forward with the introduction of the Transformer architecture. The use of Transformers, however, still requires in-depth knowledge to set up a suitable pipeline.

\begin{description} 

\item[Data Intake and Analysis.]{Users upload a single CSV for which \codename highlights column types, missing rows, and label balance. Furthermore, users can inspect class distributions and statistical analysis for numerical columns. Finally, the users can inspect the first ten rows of the dataset.}

\item[Automatic Classifier Generator.]{As illustrated in \cref{fig:classformer}, we only expose the decisions that novices are expected to understand (input columns, target labels) and run a reproducible pipeline with safe defaults, aiming to produce a working baseline on the first attempt. Model metadata are saved to eliminate train–inference mismatches and promote consistent behavior.}

\begin{figure}[t]
    \centering
    \includegraphics[width=\linewidth]{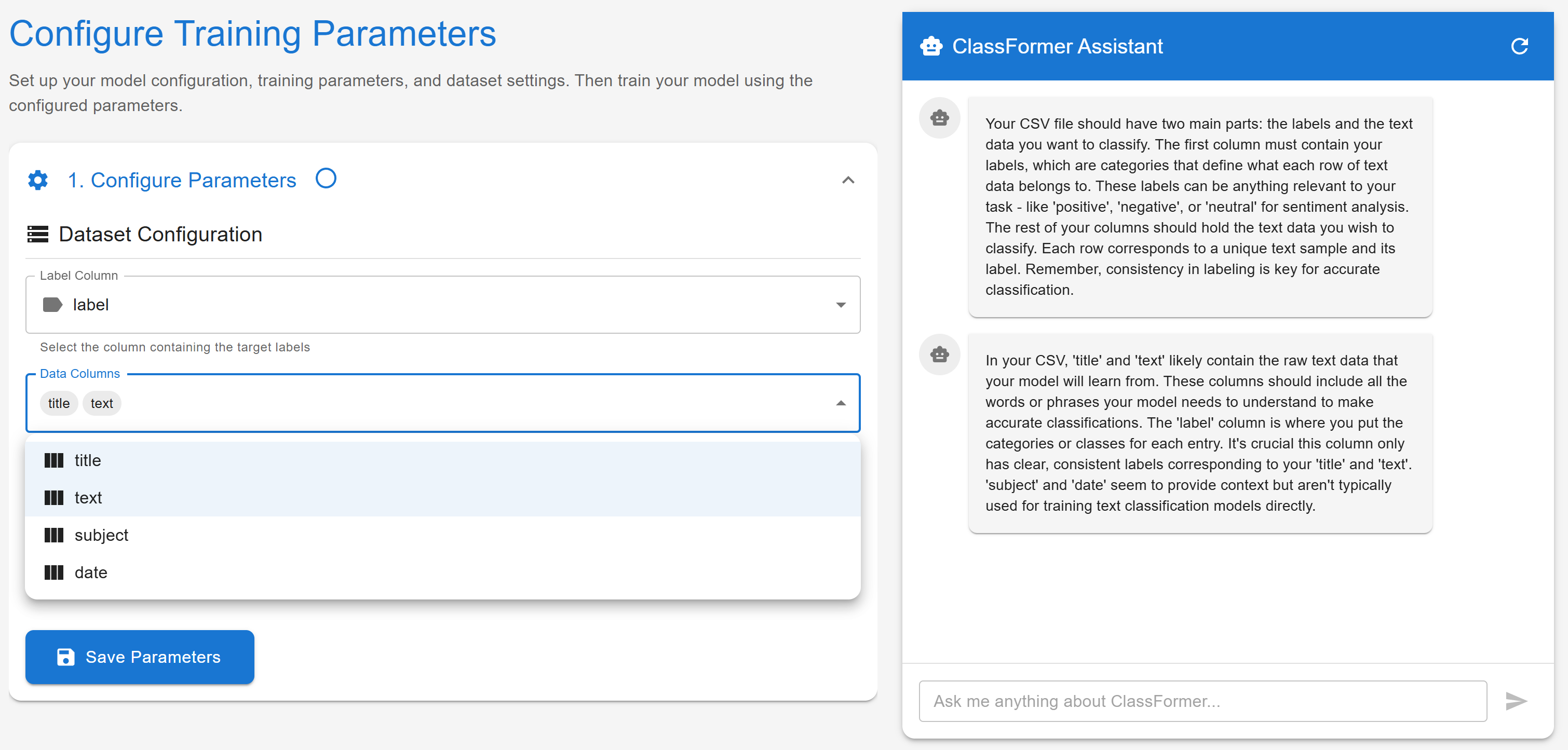}
    \caption{The \codename interface for configuring model parameters. On the left, users select which columns provide the input text and which column contains the target labels, choices that determine how the classifier is trained. On the right, an integrated contextual assistant gives real-time explanations and guidance to support novices during configuration.}
    \label{fig:classformer}
\end{figure}

\item[Cascade Classification Strategy.]{To enable novices to use advanced classification strategies, \codename integrates a one-toggle cascaded classification option. As illustrated in \cref{fig:cascade}, \codename decomposes multi-class prediction tasks into a sequence of simpler binary decisions, supporting the one-click training even with advanced classification strategies. Instead of training a single model to discriminate across all categories simultaneously, the cascade arranges multiple binary classifiers in a hierarchical structure. This sequential breakdown can reduce label imbalance and limit the risk of confusion between adjacent categories, aiming to improve recall for underrepresented classes and precision for high-confidence ones~\cite{thys_improving_2025}.}

\begin{figure}[t]
    \centering
    \includegraphics[width=\linewidth]{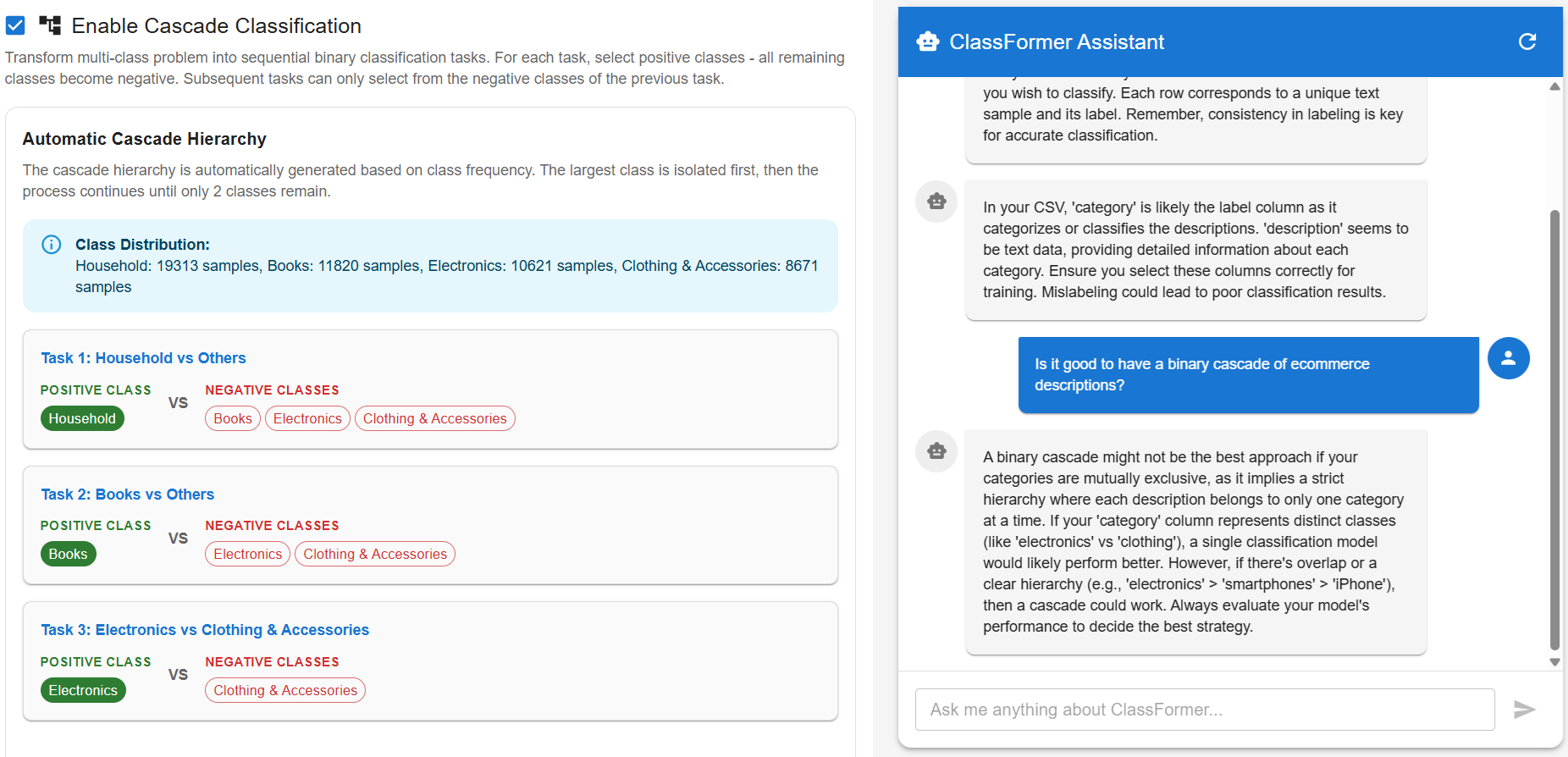}
    \caption{The \codename interface for cascade classification (left), supported by the integrated contextual assistant (right). Instead of training a single multi-class model, the system automatically builds a hierarchy of binary classifiers. This cascade is generated based on class frequency, automatically addressing label imbalance and making it easier for novices to apply advanced classification strategies without additional configuration.}
    \label{fig:cascade}
\end{figure}

\item[Inference of the Trained Model.]{The inference view uses metadata (e.g., strategy, encoders, label order), to prevent mismatches between training and inference. Outputs present both the predicted label confidence and the full class-probability distribution to support quick plausibility checks.}

\item[Conversational Assistant.]{A context-aware assistant provides simplified explanations of metrics (e.g., accuracy, F1, recall) and suggests next steps appropriate to the user's stage, aiming to reduce jargon and decision burden. We use IBM's Granite 3.3 8B\footnote{https://www.ibm.com/new/announcements/ibm-granite-3-3-speech-recognition-refined-reasoning-rag-loras} as, in our testing, it adhered best to instructions, has a context window of 128K to be able to handle larger contexts, and is small enough to run locally, as we aim to provide a secure environment for users with confidential and/or sensitive data.}

\end{description}

\section{User Testing}

\subsection{Methodology}

We conducted a study designed to evaluate how users interact with \codename across three different tasks of varying complexity. \textbf{Task 1} is a binary classification task that assesses whether users can successfully configure a simple text classifier with multiple candidate input fields. This provides a baseline of how effectively the guided dataset configuration supports novices in making informed feature and label selections. \textbf{Task 2} is a cascade classification task that examines whether participants can set up a hierarchical pipeline and allows us to test the system's ability to introduce more advanced classification strategies in an approachable manner. Finally, \textbf{Task 3} is a diagnosis task that focuses on result interpretation, testing whether the available tools help users identify performance issues due to class imbalance in the dataset. We ordered the tasks using a balanced Latin square to mitigate carry-over effects. The datasets we used are listed in \cref{app:datasets}.

The demographic information for the study was collected through a questionnaire at the beginning of the study. Participants are asked to provide their age group, highest degree obtained, and self-described gender identity. In addition, the questionnaire captured participants' prior exposure to artificial intelligence and machine learning by asking whether they had used AutoML tools before and whether they had ever trained or fine-tuned models. Finally, we asked the participants to complete the PAILQ-6 (Perceived Artificial Intelligence Literacy) questionnaire~\cite{grassini_psychometric_2024}.

To evaluate \codename, and in extension our design principles, we base our questionnaire on the study by Drozdal et al.~\cite{drozdal_trust_2020}. It targets trust and understandability through 14 items, of which two\footnote{``I understand how estimators are selected'' and ``I understand the differences between the generated models''} were removed as they are only applicable to systems where multiple model architectures can be trained and compared. Responses are collected on a 5-point Likert scale, supplemented by a binary question asking the participant whether they would deploy their trained model in the real world. Additionally, we asked the participants to complete the User Experience Questionnaire (UEQ)~\cite{laugwitz_construction_2008}.

\subsection{Results}

Our study included 24 participants (18 male, 6 female), with a majority of them in age groups 18--24 ($n=8$) and 25--34 ($n=9$). Most of them ($n=22$) indicated that they had never used AutoML tools before. To investigate differences between novice and experienced users, we created two groups: users who had previously trained ML models themselves (experienced users, $n=16$) and users who had never trained a model before (novices, $n=8$). In the experienced group, most users ($n=13$) had trained deep neural models before, while 6 users had fine-tuned models before. Four users indicated that they had trained models as part of a course, but not outside of the course. Using a Mann-Whitney U test, we found a significant effect on the Perceived Artificial Intelligence Literacy scores between the two groups. The mean ranks of the experienced group and novice group were 15.06 and 7.38, respectively ($U = 23$, $Z = 2.52$, $p < 0.05$, $r = 0.51$). The medians of participants' mean 7-point Likert ratings were 5.5 and 4.4 for the experienced group and novice group, respectively.

\subsubsection{Task Performance}

\paragraph{\textbf{Task 1.}} All participants were able to train a working classifier model to predict fake news articles. We asked the participants to rate their confidence level both for training and using a binary classification model on a 7-point Likert scale (Extremely unconfident - Extremely confident). 21 participants (87.5\%) indicated at least some level of confidence in training the model, while 3 participants (12.5\%) indicated they were extremely unconfident.

\paragraph{\textbf{Task 2.}} All participants were able to correctly train a cascade classification model to classify e-commerce descriptions. Participants were asked to determine which task in the cascade performed the worst and which tools they used to reach that conclusion. 22 participants (91.7\%) were able to correctly identify the lowest performing task. We asked participants to indicate which \codename tools they used to identify the lowest-performing task. The most-used tool was the classification report ($n=14$), closely followed by the conversational assistant ($n=13$).

\paragraph{\textbf{Task 3.}} 17 participants (70.8\%) were able to correctly identify the label imbalance in the dataset. The participants were asked to rate whether they thought they correctly identified the issue on a 5-point Likert scale (Definitely not - Definitely yes). 18 participants (75\%) reported at least some level of certainty of correctly identifying the problem. We asked participants to indicate which \codename tools they used to diagnose the issue. The most-used feature was the conversational assistant ($n=15$), closely followed by the data analysis ($n=14$) and the confusion matrix ($n=13$).

\subsubsection{User Experience, Trust, and Understandability}

\begin{figure}
    \centering
    \includegraphics[width=\linewidth]{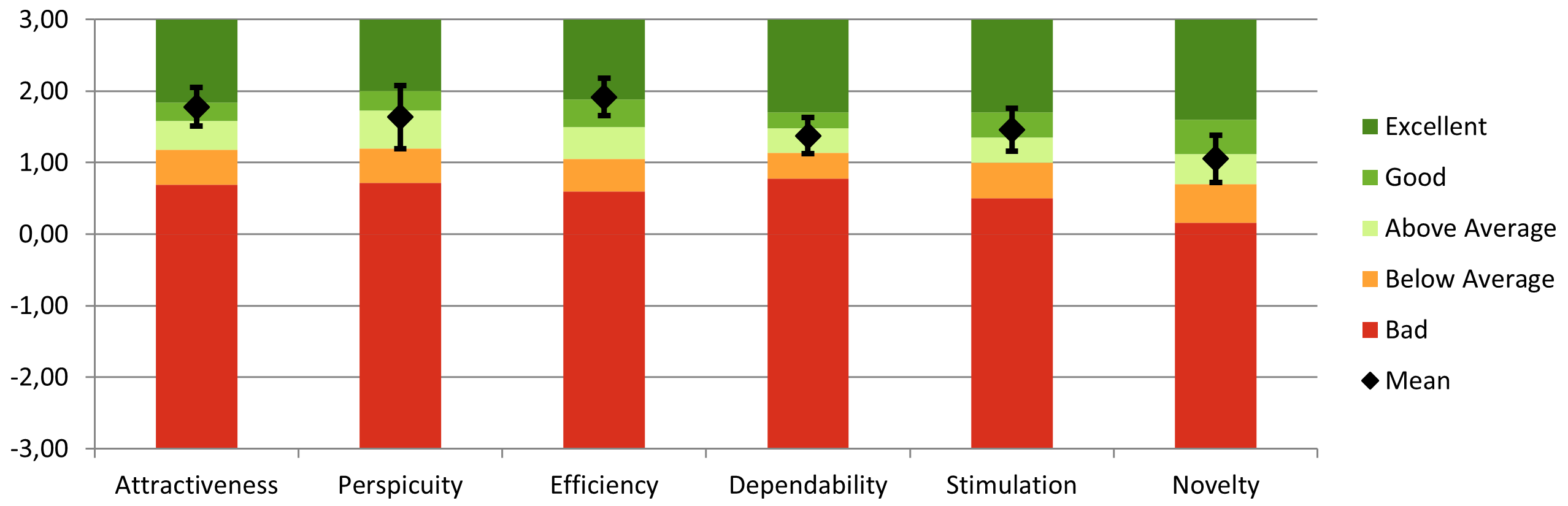}
    \caption{Distribution of UEQ responses and comparison to the UEQ benchmark. The colored background bars are reference bands from the UEQ benchmark: Bad (bottom 25\%), Below average (25-50th percentile), Above average (50-75th percentile), Good (75-90th percentile), and Excellent (top 10\%). For each scale (range -3 to +3), the black diamond and whiskers show our sample mean and 95\% CI; their position against the bands indicates the benchmark class of our product. Overall, ratings are positive, highest for Efficiency, Attractiveness, and Perspicuity; positive but more moderate for Dependability and Stimulation; and comparatively lower for Novelty.}
    \label{fig:ueq}
\end{figure}

\paragraph{\textbf{User Experience.}} Participants evaluated the system positively across all six UEQ dimensions (scale -3 to +3, full results in \cref{tab:ueq}, visualized in \cref{fig:ueq}). The highest ratings were observed for efficiency, attractiveness, and perspicuity, indicating that the tool was perceived as effective, appealing, and relatively easy to understand. Dependability and stimulation also received high ratings, suggesting that participants considered the system reliable and engaging. Novelty was evaluated somewhat lower in comparison to the other dimensions. We note, however, that the internal consistency for some scales was limited. Specifically, Cronbach's Alpha was below the commonly accepted threshold of 0.7 for Efficiency ($\alpha=0.61$) and Dependability ($\alpha=0.49$). This may indicate heterogeneous responses or that participants understood the items differently.

\begin{table}
    \centering
    \caption{UEQ results for the six standard scales (Attractiveness, Perspicuity, Efficiency, Dependability, Stimulation, Novelty) on the -3 to +3 evaluation range (higher is better). For each scale, we report the sample Mean, Std. Dev. across participants ($n = 24$), the half-width of the 95\% confidence interval, and the corresponding lower/upper bounds around the mean. We note that Cronbach's Alpha was below the commonly accepted threshold of 0.7 for Efficiency ($\alpha=0.61$) and Dependability ($\alpha=0.49$).}
    \label{tab:ueq}
    \begin{tabularx}{\textwidth}{l *{5}{>{\centering\arraybackslash}X}}
        \toprule
        \textbf{Scale} & \textbf{Mean} & \textbf{Std. Dev.} & \textbf{Confidence} & \multicolumn{2}{c}{\textbf{Confidence interval}} \\
        \cmidrule(lr){5-6}
         &  &  &  & \textbf{Lower} & \textbf{Upper} \\
        \midrule
        Attractiveness & 1.778 & 0.677 & 0.271 & 1.507 & 2.049 \\
        Perspicuity    & 1.635 & 1.096 & 0.438 & 1.197 & 2.074 \\
        Efficiency     & 1.917 & 0.658 & 0.263 & 1.653 & 2.180 \\
        Dependability  & 1.375 & 0.634 & 0.254 & 1.121 & 1.629 \\
        Stimulation    & 1.458 & 0.743 & 0.297 & 1.161 & 1.756 \\
        Novelty        & 1.052 & 0.831 & 0.332 & 0.720 & 1.384 \\
        \bottomrule
    \end{tabularx}
\end{table}

\paragraph{\textbf{Trust and Understandability.}} Using a Mann-Whitney U test, we found a significant effect of experience on the average trust and understandability scores between the two groups. The medians of participants' mean 5-point Likert ratings were 4.05 and 3.64 for the experienced group and novice group, respectively. The mean ranks of the experienced group and novice group were 14.91 and 7.69, respectively ($U = 25.5$, $Z = 2.37$, $p < 0.05$, $r = 0.48$). Next, we investigated each question from the questionnaire individually using a Mann-Whitney U test. We found a significant effect of experience between the two groups for the following questions:

\begin{itemize}
    \item \textbf{``I understand the tool.''} The median 5-point Likert ratings were 4.5 and 4 for the experienced group and novice group, respectively. The mean ranks of the experienced group and novice group were 14.75 and 8, respectively ($U = 28$, $Z = 2.43$, $p < 0.05$, $r = 0.50$).
    \item \textbf{``I understood the tool's overall process.''} The median 5-point Likert ratings were 5 and 4 for the experienced group and novice group, respectively. The mean ranks of the experienced group and novice group were 14.43 and 8.63, respectively ($U = 33$, $Z = 2.10$, $p < 0.05$, $r = 0.43$).
    \item \textbf{``I understand the data.''} The median 5-point Likert ratings were 5 and 4 for the experienced group and novice group, respectively. The mean ranks of the experienced group and novice group were 15.22 and 7.06, respectively ($U = 20.5$, $Z = 2.94$, $p < 0.05$, $r = 0.60$).
    \item \textbf{``I understand the model evaluation metrics.''}: The median 5-point Likert ratings were 5 and 2.5 for the experienced group and novice group, respectively. The mean ranks of the experienced group and novice group were 15.03 and 7.44, respectively ($U = 23.5$, $Z = 2.74$, $p < 0.05$, $r = 0.56$).
\end{itemize}

When asked whether they would deploy models trained with \codename, 17 participants (70.8\%) answered yes. The most recurring reason ($n=10$) was a variant of ``high accuracy'' or ``high F1-score,'' but experienced users would also give more detailed feedback. For example, one experienced user mentioned ``There were no fake articles mentioned as real, which indicates a low chance of the worst case scenario,'' referencing the precision and recall in the classification report. Out of the 7 participants who chose not to deploy the models, there were mainly issues about the transparency of the tool and/or the models. Participants, for example, mentioned ``I am unsure because I do not know what the model bases itself on to make a decision,'' ``No overview of how the data is processed, so not reliable to implement in production,'' and ``I can not know for sure if the test dataset that is used for the evaluation is biased towards the training data or not.'' These comments are in line with the differences we found for questions about the understanding of the data and evaluation metrics.

The low understanding of the tool and overall process can also be linked to observations made, especially during Task 2. Many users, novice and experienced, would enable the cascade classification as per the instructions, yet paid no attention to the system's explanation about what would happen, how the data would be processed, or what the final binary classifiers would look like.

\subsection{Discussion}

This study set out to understand whether a guided, end-to-end workflow can make Transformer fine-tuning accessible to novices, and how design choices shape trust and understanding. Overall, participants completed the core tasks successfully: everyone trained a working binary classifier (Task 1) and cascaded classifier (Task 2), and most participants correctly diagnosed class imbalance in the analysis task (Task 3). These outcomes suggest that \codename's training workflow and metadata-driven inference help reduce the kinds of configuration errors that commonly block novice progress. The positive UEQ ratings for efficiency, attractiveness, and perspicuity reinforce this, indicating that participants perceived the tool as effective and comprehensible in practice. 

At the same time, the results reveal a persistent gap between novice and experienced users. Experienced participants reported higher trust and understandability scores, with significant group differences on items about understanding the tool, the overall process, the data, and evaluation metrics. These differences likely reflect not only prior exposure to ML concepts but also how users interpret model feedback: those with prior experience may map metrics and visualizations to mental models more readily, whereas novices need more scaffolding to connect outputs to actionable insight. The finding that 17 of 24 would deploy their models, with high accuracy or F1-score as the most common reason, shows that many participants equate performance indicators with deployability, while the seven who hesitated emphasized the black-box nature of \codename and the trained models. Both reactions underscore the need to couple performance reporting with transparent, digestible explanations of model behavior and limits. Even so, novices did not evaluate the system negatively: their mean trust/understandability scores remained mostly positive, aligning with the broadly positive UEQ results and high task success rates. This suggests the gap reflects relative differences rather than dissatisfaction.

The conversational assistant played a central role in sense-making across tasks. Participants relied on the conversational assistant to select input features (Task 1), identify the weak cascade stage (Task 2), and diagnose imbalance (Task 3), often as much as or more than static tools like the classification report or confusion matrix. This pattern suggests that context-aware, real-time guidance can assist users by translating results into plain-language cues tied to the user's current step. However, dependence on the assistant also increases the cost of occasional hallucinations or imprecise explanations, which several participants encountered.

Defaults and automation enabled novices to quickly reach ``first model success,'' while the cascade option let all participants experience a more advanced strategy without extra configuration. Yet the uniform interface likely under-served both ends of the spectrum: novices who would benefit from additional safeguards and predictive hints, and experienced users who asked for deeper controls and richer diagnostics.

\section{Design Principles for AutoML Tools for Novices}

We translate our findings into actionable guidance for developers of AutoML tools aimed at novices. We combine the abstract pipeline, relevant theories, and insights from our study into four principles that aim to raise self-efficacy, calibrate trust, and preserve user control while still accommodating expert needs.

\begin{principle}[First-Model Success to Raise Self-Efficacy]\label{P1}

Ensure a working baseline on the first attempt so novices experience immediate success. Early mastery experiences measurably increase self-efficacy and persistence; a near-guaranteed first win is a reliable way to raise confidence for later tasks~\cite{bandura_self-efficacy_1977}. To help first-model success, we propose the following implementation guidelines:

\begin{itemize}
    \item Ship safe defaults with one-click training and make failure hard: validate inputs up front (e.g., schema, labels, missing values), choose conservative parameters, and automatically handle actions like preprocessing, label order, checkpointing, and recovery from common errors.
    \item Provide real-time feedback (e.g., ``Model trained'', ``F1-score=\dots'') combined with next-step nudges (e.g., ``Try other input features'').
    \item Combine ease of use with advanced functionality: enable advanced features (like cascade classification) through single toggles that maintain the one-click training workflow.
    \item Preserve first-model success across expertise levels: Regardless of interface adaptation or user expertise, maintain the goal of first-model success through consistent defaults and failsafe mechanisms. Dataset-agnostic strategies and advanced techniques should be deployable with one click, ensuring novices benefit from stronger baselines while experts retain inspection and override capabilities.
\end{itemize}

\end{principle}

\begin{principle}[Explanations to Create Mental Models and Appropriate Reliance]\label{P2}

Pair metrics with simplified explanations so users understand what the model did and how well it performed. Explanatory debugging improves users' mental models~\cite{kulesza_principles_2015}, and appropriate reliance requires transparency beyond raw scores~\cite{lee_trust_2004}; both can reduce over- and under-trust. Our study revealed that while 71\% of participants would deploy their models based on high accuracy scores, those who hesitated cited concerns about not knowing ``what the model bases itself on to make a decision'' and lack of ``overview of how the data is processed.'' To help create mental models and appropriate reliance, we propose the following implementation guidelines:

\begin{itemize}
    \item Add tiered explanations for key metrics. When novice users inspect a score (e.g., F1-score, precision, recall) or a visualization (e.g., confusion matrix, ROC curve), show a short tooltip describing what the metric measures, how to interpret high or low values, and one simple suggestion for improvement. When experienced users inspect performance, the system should provide more in-depth and advanced metrics.
    \item Add appropriate reliance cues next to each metric (e.g., ``High F1-score with low minority-class recall risks under-serving class Y'').
    \item Make system processes (e.g., data preprocessing, train-test splitting, validation steps) transparent and inspectable. Users need to be able to understand the system's entire process.
    \item Implement tool-augmented conversational approaches~\cite{vanbrabant_echo_2025} that enable users to interactively explore model decisions through natural dialogue. These systems combine large language models with explanatory tools, allowing users to ask questions like ``Why was this classified as X?'' or ``Show me examples where the model confuses class A and B,'' and receive context-aware, data-driven responses. Such conversational explainability improves transparency and user understanding by supporting follow-up questions and deeper exploration of model behavior.
\end{itemize}

\end{principle}

\begin{principle}[Abstractions and Context-Aware Assistance to Support the Zone of Proximal Development]\label{P3}

Offer targeted guidance through abstracted interfaces and context-aware assistance so novices operate within their zone of proximal development~\cite{wertsch_social_1979}. Abstractions and context-aware support turn opaque steps (e.g., configuration, evaluation, inference) into guided actions, helping users perform just beyond their independent ability. Our study found significant differences in understanding between novice and experienced users, with novices scoring lower on understanding the tool, overall process, data, and evaluation metrics. Additionally, many users enabled cascade classification features without engaging with explanations. To help create better abstractions and context-aware assistance, we propose the following implementation guidelines:

\begin{itemize}
    \item Implement interfaces that adapt to ML experience using brief proficiency checks to keep users in their zone of proximal development.
    \item Embed a context-aware assistant that can help check for common problems (e.g., imbalance, too little data) and suggest the next action. To interpret data, address reliability concerns through hybrid approaches that combine LLM flexibility with templated responses to reduce hallucinations (specifically of performance numbers) while maintaining natural interaction.
    \item Implement advanced, dataset-agnostic ML strategies (e.g., model multiplicity~\cite{eerlings_ai-spectra_2025}) that require no configuration (one-click training) but allow expert inspection and/or configuration.
\end{itemize}

\end{principle}

\begin{principle}[Predictability and Safeguards to Strengthen Perceived Control]\label{P4}

Make the system predictable and safe with metadata-driven UIs and strict safeguards that prevent train-inference mismatches. In the theory of planned behavior, perceived behavioral control is a key driver of intention. Clear constraints, validation, and consistency increase users' felt control~\cite{ajzen_intentions_1985}. To help create more predictability and better safeguards, we propose the following implementation guidelines:

\begin{itemize}
    \item Generate inference UIs directly from training metadata (e.g., schema, label order, input features).
    \item Provide pre-flight checks (e.g., schema, missing values, label coverage) and actionable errors with safe fallbacks.
    \item Implement validation and constraints appropriate to the user's expertise level, with novices receiving more protective guardrails and experienced users having more flexibility.
    
\end{itemize}

\end{principle}

\section{Conclusion}

This paper contributes an \textbf{abstract end-to-end pipeline covering data intake, configuration, training, evaluation, and inference}. The emphasis of this workflow is on reliability for novices and user understanding rather than algorithmic optimization for peak performance. We used a \textbf{prototype implementation} to examine the abstract pipeline in practice. In a \textbf{24-participant study}, all participants successfully trained a binary and cascaded classifier. \textbf{User-experience ratings were positive}, and 17 participants reported they would deploy their models. However, \textbf{experienced users reported higher trust and understanding} than novices, and several participants raised transparency concerns. 

Based on these findings and relevant theories, we propose a set of \textbf{four design principles for AutoML tools for novices}: (\cref{P1}) first-model success, (\cref{P2}) explanations that support correct mental models and appropriate reliance, (\cref{P3}) abstractions with context-aware assistance, and (\cref{P4}) predictability through safeguards and metadata. Based on these design principles, future AutoML tools could (i) combine LLM-driven guidance with templated summaries to reduce hallucinations and maintain step awareness, (ii) implement expertise-adaptive scaffolding that preserves first-run defaults while progressively revealing controls, and (iii) develop accessible explainability that links metrics and visualizations to concise, actionable narratives.

\begin{credits}
\subsubsection{\ackname}
This work was supported by the Special Research Fund (BOF) of Hasselt University (BOF24OWB28). This research was made possible with support from the MAXVR-INFRA project, a scalable and flexible infrastructure that facilitates the transition to digital-physical work environments. The MAXVR-INFRA project is funded by the European Union - NextGenerationEU and the Flemish Government.

\subsubsection{\discintname}
The authors have no competing interests to declare that are relevant to the content of this work.
\end{credits}

\bibliographystyle{splncs04}
\bibliography{references}

\appendix

\section{Datasets}\label{app:datasets}

\noindent\href{https://www.kaggle.com/datasets/clmentbisaillon/fake-and-real-news-dataset}{Task 1: Fake News Classification with Multiple Input Columns}

\noindent\href{https://doi.org/10.5281/zenodo.3355822}{Task 2: Cascaded Product Category Classification}

\noindent\href{https://www.kaggle.com/datasets/mahmoudelhemaly/students-grading-dataset}{Task 3: Diagnosis of Imbalance in Grade Prediction}

\end{document}